\documentclass[aps]{revtex4-1}
\usepackage{graphicx}
\begin{document}
\title{Phase diagram of UGe$_2$.\\
{\it Whether there are quantum phase transitions ?}}

\author{V.P.Mineev}

\affiliation{Commissariat \`a l'Energie Atomique,
INAC/SPSMS, 38054 Grenoble, France}


\begin{abstract}
The phase diagram of several itinerant ferromagnets reveals the common feature. The phase transition temperature decreases with pressure increase and reaches zero value at some critical pressure $P_c$ such that  at low enough temperatures one can expect  critical behavior specific for quantum phase transition.  It is not the case, however. Being the second order 
at ambient pressure   the transition from paramagnetic to ferromagnetic state  at high pressures - low temperatures is transformed to the discontinuous jump. 

We discuss
the magneto-elastic mechanism of development of the first order type instability at the phase 
transition to  the ferromagnet state  in strongly anisotropic ferromagnet  UGe$_2$. 
Using the parameters characterizing the  properties of UGe$_2$  we argue the effectiveness of this mechanism  transforming the very weak  first order type transition to the really observable one.
\end{abstract}
\maketitle
\section{Introduction}

The pressure-temperature phase diagrams of several weak ferromagnets exhibit similarity. The  transition from the paramagnetic to the ferromagnetic states at ambient pressure occurs  by means of the second order phase transition. The phase transition temperature decreases with pressure increase such that it reaches zero value at some critical pressure $P_c$.  At some  pressure interval below  $P_c$ the ordered ferromagnetic moment disappears discontinuously. Thus at high pressures and low temperatures  the ferromagnetic and paramagnetic states are divided by the   first order type transition whereas at higher temperatures and  lower pressures this transition is of the second order. Such type of behavior is typical for
MnSi  [1-4], itinerant ferromagnet-superconductor UGe$_2$ 
[5,6] (see Fig.1), ZrZn$_2$ [7]. The same behavior has been established  in the ferromagnetic compounds Co(Si$_{1-x}$Se$_x$)$_2$ 
[8]
and (Sr$_{1-x}$Ca$_x$)RuO$_3$ [4] where the role of governing parameter plays 
the concentration of Se and Ca correspondingly.
Also, there was demonstrated clear evidence  for the first order nature of the ferromagnetic transitions in typical ferromagnets like Ni, Fe and Co [9].
\begin{figure}[h]
\includegraphics[width=3in]{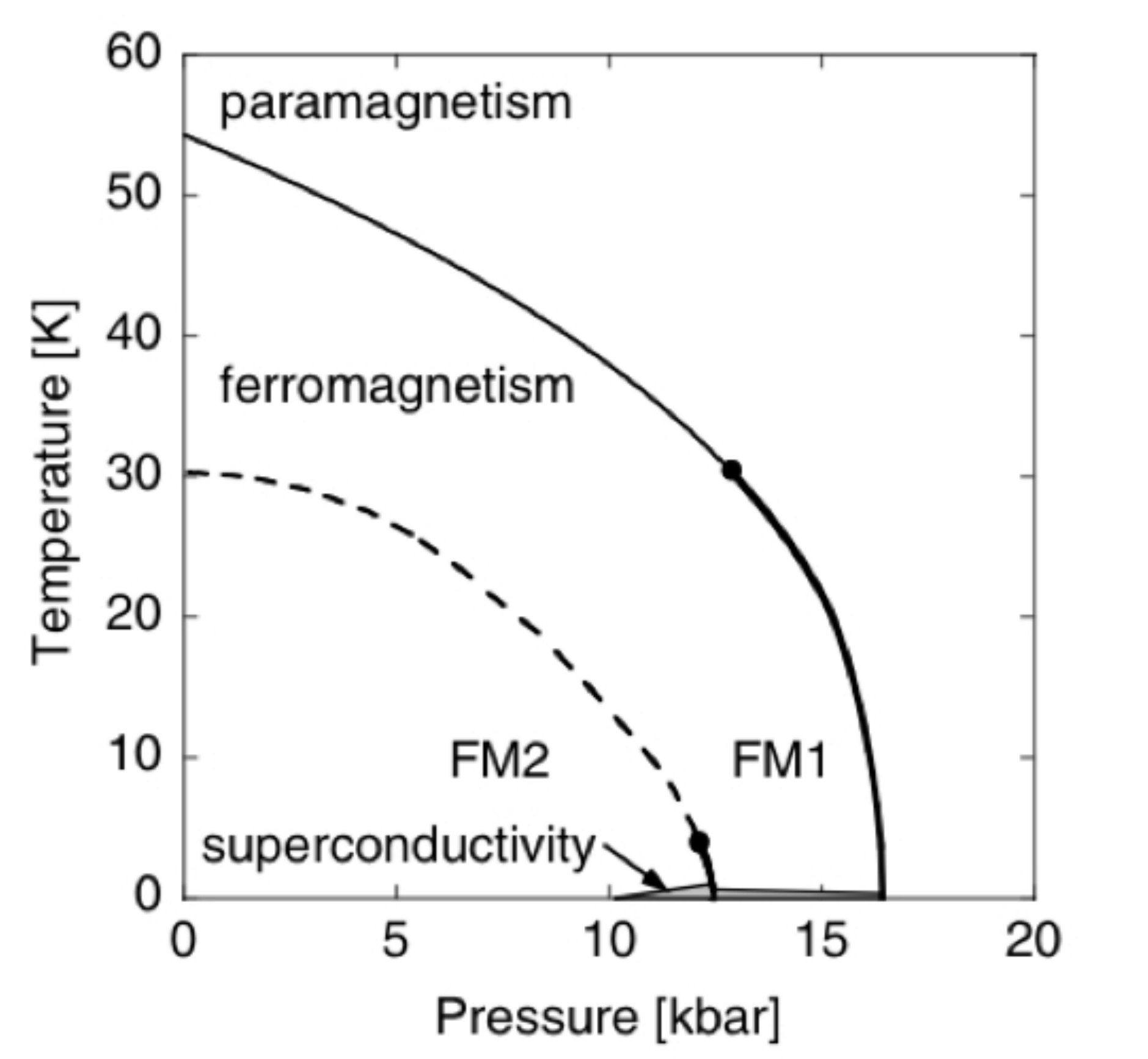}\hspace{2pc}%
\begin{minipage}[b]{14pc}\caption{\label{label}
The schematic phase \\diagram of UGe$_2$ Ref. [10]. Thick lines denote first order transitions and fine lines second order transitions. The dashed line is a crossover. Dots mark the positions of critical (tricritical) points. 
}
\end{minipage}
\end{figure}

Here we discuss the magneto-elastic mechanism  of development of the first order type instability.
Actually the mean field treatment of the magneto-elastic  mechanism   has been put forward  
in the paper [11] 
 where it was demonstrated that the change of transition character from the second to the first order
  takes place at strong enough steepness  of the exchange interaction dependence on interatomic distance and large compressibility. It can be considered in frame of the Landau  theory of the phase transition. Namely,  in neglect the shear deformation the free energy density near  the phase transition to the  Ising type ferromagnet has the following form
\begin{equation}
F=\alpha_0(T-T_c)M^2+\beta M^4+\frac{K}{2}\varepsilon^2-q\varepsilon M^2.
\end{equation}
Here, $M$ is the magnetization density, $\varepsilon$ is the relative volume change, $K$ is the bulk modulus.
The coefficient $q$ is related to the Curie temperature pressure dependence as
\begin{equation}
q=\alpha_0\frac{d T_c}{d\varepsilon}=-\alpha_0K\frac{d T_c}{d P}.
\end{equation}
In  stress absence $
\frac{\partial F}{\partial \varepsilon}=0$, the deformation is determined by square
of magnetization $\varepsilon=\frac{q}{K}M^2$ that yields
\begin{equation}
F=\alpha M^2+\left ( \beta-\frac{q^2}{2K}   \right )M^4.
\end{equation}
Hence, at $
\frac{q^2}{2K} >\beta$ the phase transition changes its character from the second  to the first order.
This inequality can be rewritten through the measurable parameters as
\begin{equation}
\frac{K\Delta C}{T_c}\left (\frac{d T_c}{d P}\right)^2>1,
\end{equation}
where we used  the formula $\Delta C=\frac{\alpha_0^2}{2\beta}T_c$ for the specific heat jump at phase transition of the second order.

The magneto-elastic interaction also produces
another general mechanism
 for instability of second order phase transition 
 toward to the discontinuous  formation of ferromagnetic state from  the paramagnetic one.  
 For the first time it was pointed out  by O. K. Rice [12] 
who has demonstrated that at small enough distance from the volume dependent critical temperature $T_c(V)$, where the specific heat $C_{fl}(\tau) \sim \tau^{-\alpha},$ $\tau = \frac{T}{T_c(V )} -1$, tends to infinity due to the critical fluctuations,
the system bulk modulus $K= -V \frac{\partial P}{\partial V}=V\frac{\partial^2 FV}{\partial V^2}$, expressed through the free
energy density $F = F_0 + F_{fl}, ~~F_{fl}\sim -T_c\tau^{2-\alpha} $ starts to be negative
\begin{equation}	
K=K_0-A\frac{C_{fl}(\tau)V^2}{T_c}\left (\frac{\partial T_c}{\partial V}\right)^2 
=K_0-\left.AK_0^2\frac{C_{fl}(\tau)}{T_c}\left (\frac{\partial T_c}{\partial P}\right)^2 \right |_{\tau\to 0}< 0~,
\end{equation}	
that	contradicts	to thermodynamic stability of the system. 
 In reality, before there will be reached the temperature corresponding to $K=0$
 the system undergoes  the first order transition, such that to jump over the instability region directly in the ferromagnetic state with finite magnetization and related to it striction deformation. This transition is similar to the jump over the region with $\partial P/\partial V >0$  on the van der Waals isotherm at the liquid-gas transition. 
 
 The condition of the first order instability (5)  can be  written in similar to Eqn.(4) form
 \begin{equation}	
\frac{K_0C_{fl}(\tau)}{T_c}\left (\frac{\partial T_c}{\partial P}\right)^2 >1.
\end{equation}	
However, unlike to Eq. (4) this formula demonstrates that the  first order instability is inevitable due to infinite increase of fluctuation specific heat.
 Thus, if  in the system with the fixed volume the phase transition is of the second order with the infinite increase of specific heat then 
  the effect of finite compressibility under assumption that the critical temperature is the volume dependent  parameter transforms it into the phase transition of the first order.  
In reality,  the striction interaction can change the shape of the free energy singularity in respect to its form in incompressible case. More elaborate treatment [13] taking into account this effect leads to the following condition of the first order instability 
$
\frac{1}{T_c}\frac{4\mu K}{3K+4\mu}f''(x)\left (\frac{\partial T_c}{\partial P}\right )^2>1.
$
Here the function $f(x)$ determines the fluctuation part of free energy $F=-T_cf\left (\frac{T-T_c}{T_c}\right)$, $\mu$ is the shear modulus. Usually, the left hand side in Eqn. (6) is quite small
and the transition of the first order occurs at temperature $T^\star$ close to the critical temperature  where fluctuation specific heat is large enough. It means that the  temperature difference $T^\star-T_c$ is smaller than the critical temperature $T_c$ by many orders. The latent  heat at this transition 
$
q\approx C_{fl}(T^\star-T_c)
$
proves to be extremely small. So, the first order phase transition is practically indistinguishable from the second order one and called weak first order phase transition or the phase transition of the first order closed to the second order. 

According to Eqs  (4), (6)  the magneto-elastic mechanism  effectively leads to the first order transition
when the critical temperature is strongly pressure dependent. This is the case in all mentioned above materials.  To check the criteria (4), (6) one must calculate the mean field jump and fluctuation part of the specific heat near the Curie temperature for each given material. 
To be concrete, here, I'll do these calculations for
UGe$_2$ characterized by strong magnetic anisotropy and by
 the precipitous drop of the critical temperature at pressure increase near 14-15 kbar [14].

\section{The specific heat near the Curie temperature}

UGe$_2$ is orthorhombic crystal with ferromagnetic order at ambient pressure found below $T_c=53~K$. 
Magnetic measurements reveal a very strong magnetocrystalline anisotropy [15] with ${\bf a}$ being the easy axis.  We shall denote it as $z$ direction.
The free energy  of strongly anisotropic ferromagnet can be written in terms of one component scalar order parameter corresponding to magnetization density  $M_z({\bf r})$ along $z$ axis. In that follows we shall omit the order parameter index z.
\begin{equation}
{\cal F}=\int d^3{\bf r}\left\{\alpha M^2+ \beta M^4+\gamma_{ij}\nabla_iM\nabla_jM
-\frac{1}{2}\frac{\partial^2M({\bf r})}{\partial z^2}\int\frac{M({\bf r}')d^3{\bf r}'}{|{\bf r}-{\bf r}'|}\right\}
\end{equation}
Here, $\alpha=\alpha_0(T-T_c) $, the gradient terms are written taking into account the orthorhombic anisotropy
$
\gamma_{ij} = \left(\begin{array}{ccc} \gamma_{xx} & 0 & 0\\
0 & \gamma_{yy} & 0 \\
0 & 0 & \gamma_{yy}
\end{array} \right),
$
where the $x, y, z$ are directions of the spin axes pinned to $b, c, a$
crystallographic directions correspondingly. The last nonlocal term in Eq. (7) corresponds to magnetostatic energy [16,17] $-{\bf M}{\bf H}-H^2/8\pi$, where internal magnetic field ${\bf H}$ expressed in terms of magnetization density by means of Maxwell equations $rot{\bf H}=0$ and $div({\bf H}+4\pi{\bf M})=0$.
We shall use the following estimations for the coefficients in the Landau free energy functional
\begin{eqnarray}
&\alpha_0&=\frac{1}{m^2n},\\
&\beta&=\frac{T_c}{2(m^2n)^2n},\\
\gamma_x\approx&\gamma_y&\approx\gamma_z\approx\frac{T_ca^2}{m^2n}.
\label{3}
\end{eqnarray}

Here, $m=1.4\mu_B$ is the magnetic moment per uranium atom at zero temperature [18],
$n=a^{-3}$ is the density of uranium atoms, which can be approximately taken equal to inverse cube of the nearest-neighbor uranium atoms separation $a=3.85$ Angstrom [19].

The  mean field magnetization and the jump of specific heat are
\begin{eqnarray}
&M^2&=-\frac{\alpha}{2\beta}=(mn)^2\frac{T_c-T}{T_c}\\
&\Delta C&=\frac{T_c\alpha_0^2}{2\beta}=n.
\end{eqnarray}
The experimentally found specific heat jump $\Delta C_{exp}\approx 10\frac{J}{mol K}\approx 1$ per uranium atom [19] is in remarkable correspondence with Eq.(12).

The effective Hamiltonian of noninteracting field of the order parameter fluctuations   is given by
\begin{equation}
H_0=\sum_{\bf k}\left (\alpha +\gamma_{ij}k_ik_j+2\pi k_z^2/k^2\right)
M_{\bf k}M_{-{\bf k}},
\end{equation}
where $M_{\bf k}=\int M({\bf r})e^{-i{\bf k}{\bf r}}d^3{\bf r}$.
The corresponding  free energy and the specific heat are [20]
\begin{equation}
{\cal F}_{fl}=-\frac{T}{2}\sum_{\bf k}\ln\frac{\pi T}{\alpha +\gamma_{ij}k_ik_j+2\pi k_z^2/k^2},
\end{equation}
\begin{equation}
C_{fl0}=\frac{T^2\alpha_0^2}{2(2\pi)^3}\int\frac{dk_xdk_ydk_z}{[\alpha+2\pi \hat k_z^2+\gamma_{ij}k_ik_j]^2}. 
\end{equation}
Proceeding to spherical coordinates  and performing integration over modulus $k$ we come to
\begin{equation}
C_{fl0}=\frac{T^2\alpha_0^2}{32\pi^2}\int_0^1d\zeta\int_0^{2\pi}\frac{d\varphi}{(\alpha+2\pi \zeta^2)^{1/2}(\gamma_{\perp}+\zeta^2(\gamma_z-\gamma_{\perp}))^{3/2}}. 
\label{C}
\end{equation}
Here, $\gamma_{\perp}(\varphi)=\gamma_{x}\cos^2\varphi+\gamma_{y}\sin^2\varphi$.
At critical temperature $\alpha=0$ and the integral diverges. 
Hence, performing integration over $\zeta$ with logarithmic accuracy we obtain
\begin{equation}
C_{fl0}=\frac{T_c^2\alpha_0^2}{32\pi\sqrt{2\pi}\gamma^{3/2}}\ln\frac{\alpha}{2\pi}\approx\frac{n}{32\pi}\sqrt{\frac{T_c}{2\pi m^2n}}\ln\frac{2\pi m^2n}{T-T_c},
\label{CC}
\end{equation}
where $$\frac{1}{\gamma^{3/2}}=\frac{1}{2\pi}\int_0^{2\pi}\frac{d\varphi}{\gamma_\perp^{3/2}(\varphi)}.$$

The used condition $\alpha \ll 2\pi$ at $T_c=10K$ is realized at
\begin{equation}
\frac{T-T_c}{T_c}<\frac{2\pi m^2n}{T_c}\approx 0.015.
\label{T}
\end{equation}
In view of roughness of the parameter estimation the region of logarithmic increase of specific heat
can be in fact broader.

The  calculation taking into account the interaction of fluctuations has been performed by Larkin and Khmelnitskii [21].
In our notations
the expression  for the fluctuation specific heat at const pressure obtained in this paper is
\begin{equation}
C_{fl}=\frac{3^{1/3}T_c^2\alpha_0^2}{16\pi\gamma_{LK}^{2/3}\gamma^{3/2}}
\left (\ln\frac{\alpha}{2\pi}\right )^{1/3}
\label{Khm}
\end{equation}
Here $\gamma_{LK}=\frac{3T_c\beta}{\sqrt{32\pi}\gamma^{3/2}}$ is the effective constant of interaction.  Using the Eqs. (8)-(10) one can rewrite Eq. (\ref{Khm}) as
\begin{equation}
C_{fl}\approx\frac{n}{10}\left (\frac{T_c}{2\pi m^2n} \right)^{1/6}\left (\ln\frac{2\pi m^2n}{T-T_c}\right )^{1/3}.
\label{Khme}
\end{equation}

So, we see that the order parameter fluctuations give rise the increase of specific heat near the critical point. The power of the logarithm $(\ln\frac{\alpha}{2\pi})^{1/3}$ is quite slow function slightly exceeding unity, 
hence in the temperature region given by inequality (18)
one may estimate the fluctuation specific heat as 
\begin{equation}
C_{fl}>\frac{n}{5}.
\label{Khmel}
\end{equation}
We see that the fluctuation specific heat is smaller than the mean field jump given by Eqn. (12). Hence  to check the first order phase transition instability in UGe$_2$ one must to proceed with criterium (4).

\section{Instability of the second order phase transition}

The Curie temperature in UGe$_2$ falls monotonically with increasing pressure from 53 K at ambient pressure and drops precipitously above 15 Kbar [14]. The average value of the critical temperature derivative can be estimated as 
\begin{equation}
 \frac{\partial T_c}{\partial P}\approx\frac{40~Kelvin}{14~kbar}=4\times 10^{-25}~cm^3
 \label{est}
\end{equation} 
 For the bulk modulus we have 
 \begin{equation}
 K= \rho c^2\approx 10^{11}erg/cm^3,
 \end{equation}
 where we have substituted  typical sound velocity  $c\approx10^5~cm/sec$ and used known [22] density value $\rho=10.26~g/cm^3$. Thus,  
 we have for the combination Eq.(4)
 
\begin{equation}
\frac{Kn}{T_c}
\left (\frac{\partial T_c}{\partial P}\right )^2=0.2~.
\label{LPi}
\end{equation}
At $T\approx10K$  the pressure derivative of the critical temperature is much higher (and its square is even more higher) than its average value given by Eq. (22). So, we come to conclusion 
that at critical temperature of the order  10 K  the criterium (4) is fulfilled and the phase transition of the second order turns into
 the first order one.

\section{Conclusion}
The magneto-elastic interaction provides development of the first order instability at the phase transition to the ordered state in a ferromagnet. However,  actual temperature interval of this  instability development is negligibly small and the first order transition looks almost indistinguishable from the second order one.
The particular feature of anisotropic ferromagnet UGe$_2$ is the precipitous drop of the Curie temperature as the function of pressure near 14-15 kbar. Due to this property
at about these pressures the second order phase transition (or very weak transition of the first order)  to ferromagnet state turns into the real first order type transition.  

At low temperatures according to the Nernst law and the Clausius-Clapeyron relation
\begin{equation}
\frac{d T_c}{d P}=\left.\frac{v_1-v_2}{s_1-s_2}\right|_{T\to0}\rightarrow\infty
\end{equation}
 the drop of  transition temperature with pressure  begins to be  infinitely fast. It means that weak first order transition has the tendency to be stronger and stronger  as temperature decreases.  Hence, the effect of magneto-elastic interaction or, more generally,  of the order parameter interaction with elastic degrees of freedom 
at arbitrary type of ordering raises  the doubts upon the existence of quantum critical phenomena.

\subsection{Acknowledgments}
This work was partly supported by the grant SINUS of  Agence Nationale de la Recherche.

\smallskip

\end{document}